# Leading Vacuum–Polarization Contributions to the Relation Between Pole and Running Masses

**Kostas Philippides** and **Alberto Sirlin**

*Department of Physics, New York University*
*4 Washington Place*
*New York, NY 10003, USA*

**Abstract**

The vacuum–polarization contributions of $\mathcal{O}(b^{n-1}\alpha_s^n)$ to the relation between the pole–mass $M$ of a quark and the $\overline{\rm MS}$ parameter $\hat{m}(M)$ are evaluated by a straightforward method. They are found to approximate very well the exact answer, known through $\mathcal{O}(\alpha_s^2)$, thus providing a simple physical interpretation. Results are also given for the cases when the vacuum–polarization contributions are defined by the pinch technique prescription and specific background field gauges. Assuming that the terms $n \geq 3$ are also dominant, we evaluate $M_t/\hat{m}_t(M_t)$, compare the results with those of optimization methods, briefly discuss $M_b/\hat{m}_b(M_b)$, and estimate the irreducible errors in the perturbative series. Implications for the electroweak amplitude $\Delta\rho$ are emphasized. An update of the QCD corrections to this amplitude, including an estimate of the theoretical error, is given.

## 1 Leading Vacuum-Polarization Contributions to $M/\hat{m}(M)$

The relation between the pole and running quark masses is a question of considerable practical and conceptual interest. For example, in electroweak physics it has been recently observed [1–2] that the QCD corrections to the dominant $m_t$–dependent contributions are very small through $O(\alpha_s^2)$ when the amplitudes are expressed in terms of $\hat{m}_t(M_t)$ (henceforth $M$ and $\hat{m}$ denote the pole and the $\overline{\rm MS}$ running masses, respectively). On the other hand, the pole mass is the concept most closely related to mass measurements based on kinematical considerations [3]. The pole mass is also employed in many calculations of the semileptonic and Higss–boson decays, and in the evaluation of $\hat{m}(M)$ for the $b$ and $c$ quarks [4]. Recently, there has also been considerable interest in studies of contributions to the pole mass associated with renormalons, leading to the conclusion that there is an irreducible uncertainty $\sim \Lambda_{\rm QCD}$ in the perturbative definition of this concept [5–11].

For the ratio $M/\hat{m}(M)$ there is an important exact perturbative expansion through $O(\alpha_s^2)$, due to Gray, Broadhurst, Grafe, and Schilcher [12] :

$$\frac{M}{\hat{m}(M)} = 1 + \frac{4}{3}a(M) + Ka^2(M) \ , \tag{1}$$

where $a(\mu) \equiv \alpha_s(\mu)/\pi$ and $K$ is given by [12,13]

$$K = 16.0065 - n_f \ 1.0414 + 0.1036 + \frac{4}{3}\sum_{i=1}^{n_f} \Delta\left(\frac{M_i}{M}\right) \ . \tag{2}$$

The first term in Eq.(2) corresponds to the quenched approximation, while the second and the third are the vacuum polarization contributions of $n_f$ massless quarks and the quark of mass $M$, respectively. The $\Delta$ terms represent the mass corrections associated with $n_f$ "light quarks" with masses $M_i < M$ ($\Delta(r)/r$ is roughly constant, being equal to $\frac{\pi^2}{8}$ at $r=0$, $\approx 1.04$ at $r \approx 0.3$ and $\frac{\pi^2-3}{8}$ at $r=1$). Numerically, $K$ is quite large : 10.96 for the top, 12.5 for the bottom, and 13.2 for the charm quarks, respectively.

Recently [1–2], Eq.(1) in the top quark case has been studied using the BLM [14], PMS

[15], and FAC [16] optimization methods, neglecting the small $\Delta$ terms. In particular, for the BLM procedure, one has

$$\frac{M_t}{\hat{m}_t(M_t)} = 1 + \frac{4}{3}a(\mu_t^*) - 1.072\ a^2(\mu_t^*)\ , \tag{3}$$

with $\mu_t^* = 0.0960 M_t$. When applied to Eq.(1) in the top quark case, the three optimization methods give remarkably close answers, but at the same time they strongly suggest that the higher order terms in Eq.(1) have large and increasing coefficients, $\sim 10^2$ and $\sim 10^3$ in the $O(a^3)$ and $O(a^4)$ contributions, respectively. The most rigorous way to settle this issue would be to evaluate exactly the next term in Eq.(1). However, because of the difficulty of exact multi–loop calculations, prospects for this appear at present to be remote[1]. The question naturally arises as to whether there is a simple way to understand large and potentially dominant contributions to Eq.(1) occurring in $O(a^2)$ and higher. Following the QED example, a natural idea is to incorporate the large effects of one–loop vacuum–polarization bubbles (see Fig.1), which may be interpreted as a renormalization of $(g_s^0)^2$, the bare strong coupling constant present in the basic one–loop contribution. However, unlike the QED case, the concept of vacuum–polarization is ill–defined in non–abelian theories. The reason is that the familiar self–energy loops, associated with vacuum polarization, are gauge dependent. To underscore the magnitude of this problem, we recall that in the $R_\xi$ gauges the logarithmic contributions to the one–loop diagrams do not coincide with those associated with the running of $\alpha_s$. A significant improvement can be achieved by identifying the one–loop diagrams in Fig.1 with the pinch technique (PT) self–energy $\widehat{\Pi}(k^2)$ [17]. The PT is an algorithm that automatically re–arranges one–loop $R_\xi$ amplitudes into $\xi$–independent contributions endowed with desirable theoretical features. In particular, the logarithms associated with the running of $\alpha_s$, as well as the full fermionic contribution, are automatically included in $\widehat{\Pi}(k^2)$. Defining the self–energy tensor as $-i$ times the associated Feynman diagrams, writing

[1] Private communication from D.J. Broadhurst.

$\widehat{\Pi}_{\mu\nu}(k) = (k^2 g_{\mu\nu} - k_\mu k_\nu)\,\widehat{\Pi}(k^2)$, and evaluating the PT self–energy $\widehat{\Pi}(k^2)$ in dimensional regularization, we have

$$\widehat{\Pi}(k^2) = 6(g_s^0)^2(-k^2)^{-\epsilon}(4\pi)^{\epsilon-2}\Gamma(\epsilon)\mathrm{B}(2-\epsilon,2-\epsilon)\left[\frac{11-7\epsilon}{1-\epsilon} - \frac{2}{3}n_f\right] , \tag{4}$$

where $k$ is the external gluon momentum, $\epsilon = 2 - D/2$, $D$ is the dimension of space–time, and B is Euler's Beta function. The first term in Eq.(4) is the $\xi$–independent gluonic contribution obtained by adding the pinch parts to the usual self–energy, while the second represents the fermionic terms ($n_f$ is the number of light quarks regarded as massless). Alternatively, Eq(4) can be written in the form

$$\widehat{\Pi}(k^2) = 6(g_s^0)^2(-k^2)^{-\epsilon}(4\pi)^{\epsilon-2}\Gamma(\epsilon)\mathrm{B}(2-\epsilon,2-\epsilon)b\left[1 + \frac{4}{b}\frac{\epsilon}{1-\epsilon}\right] , \tag{5a}$$

where $b = 11 - 2n_f/3$ is the first coefficient of the $\mathrm{SU}(3)_c$ $\beta$–function. In the $\epsilon \to 0$ limit, using Eq.(8) and implementing the $\overline{\mathrm{MS}}$ renormalization, this becomes

$$\widehat{\Pi}(k^2)\Big|^{\overline{\mathrm{MS}}} = \frac{\alpha_s(\mu)}{4\pi} b\left[\ln(-\frac{\mu^2}{k^2}) + \frac{5}{3} + \frac{4}{b}\right] . \tag{5b}$$

The mechanism whereby pinching on the relevant Feynman diagrams generates the lowest–order $\widehat{\Pi}(k^2)$ contribution to the quark self–energy has been shown in Ref.[18]. It has been recently pointed out [19] that the PT self–energy coincides with the background–field–gauge (BFG) self–energy $\Pi(k^2,\xi_Q)_{BFG}$ evaluated for $\xi_Q = 1$, where $\xi_Q$ is the gauge parameter associated with the quantum loops. At the one–loop level, and for arbitrary $\xi_Q$, the latter is given by[2]

$$\begin{aligned}\Pi(k^2,\xi_Q)_{BFG} = \;& 6(g_s^0)^2(-k^2)^{-\epsilon}(4\pi)^{\epsilon-2}\Gamma(\epsilon)\mathrm{B}(2-\epsilon,2-\epsilon)\\ & \times\, b\,\{1 + (4\epsilon/b(1-\epsilon))\,[1 - (1-\xi_Q)(7+\xi_Q)(3-2\epsilon)/16]\} .\end{aligned} \tag{6a}$$

In the $\epsilon \to 0$ limit, employing again Eq.(8) and the $\overline{\mathrm{MS}}$ renormalization, Eq.(6a) reduces to [20]

[2] An expression equivalent to Eq.(6a) has been also derived by G. Weiglein (private communication).

$$\Pi(k^2,\xi_Q)|^{\overline{\rm MS}}_{BFG} = \frac{\alpha_s(\mu)}{4\pi} b\left\{\ln(-\frac{\mu^2}{k^2}) + \frac{5}{3} + \frac{4}{b}\left[1 - \frac{3}{16}(1-\xi_Q)(7+\xi_Q)\right]\right\} . \tag{6b}$$

Eqs.(5) and (6) indeed coincide for $\xi_Q = 1$, as well as for $\xi_Q = -7$ [20]. From Eq.(6b) we see that the BFG self–energy also automatically includes the logarithms associated with the running of $\alpha_s$, but the non–logarithmic terms depend now on the gauge parameter $\xi_Q$. Thus, the concept of vacuum–polarization is also undefined in the BFG approach. A unique one–loop answer emerges, however, if one applies the PT within the BFG amplitudes [20], in which case one obtains once more $\widehat{\Pi}(k^2)$.

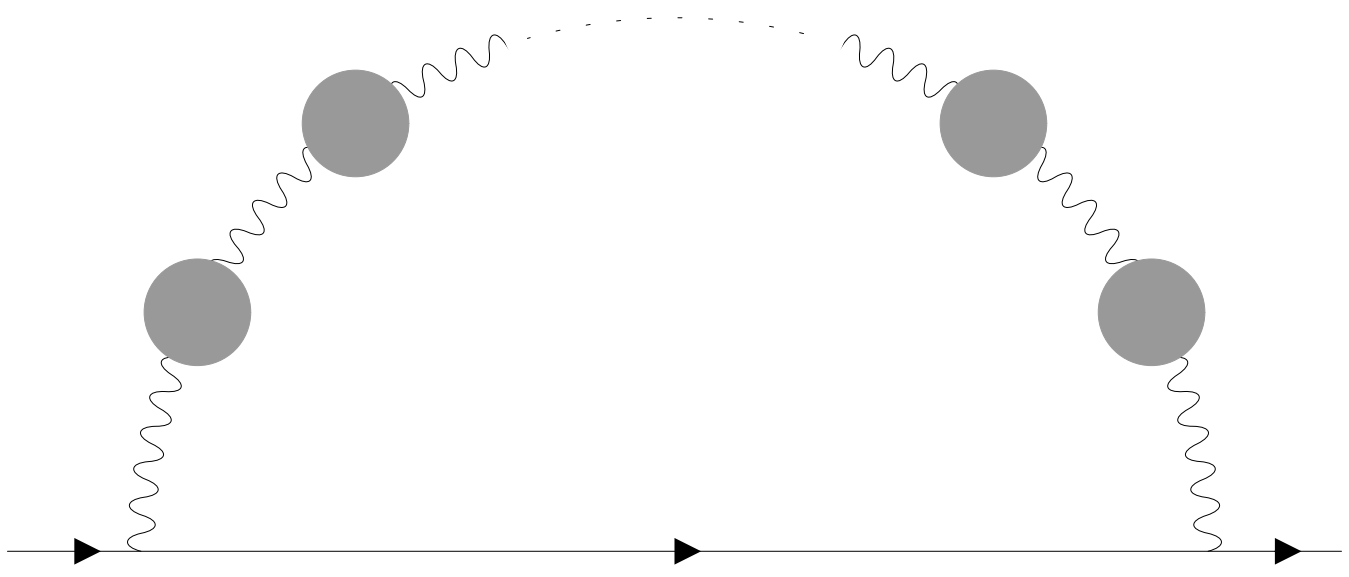

**Figure 1**

Vacuum polarization contributions to the quark self–energy , involving a chain of one–loop diagrams. Each bubble can be identified with the PT or BFG one–loop gluon self–energy (see text).

In the first part of our analysis we retain only contributions of $\mathcal{O}(b^{n-1}{\alpha_s}^n(\mu))$ $(n = 1, 2, ...)$ to $M/\hat{m}(\mu)$, where the expansion is mathematically defined so that there is no residual $n_f-$dependence in the corresponding cofactors. The ab-initio rationale is that $b$ is large and that these contributions are uniquely associated with the chains of one–loop bubbles depicted in Fig.1, while terms of $\mathcal{O}(b^l {\alpha_s}^n)$ $(l \le n-2)$ also arise from other diagrams. A more powerful a–posteriori argument is that, as we will explicitly show, the retained contributions approximate very well, through $\mathcal{O}({\alpha_s}^2)$, the exact result of Eq.(1). Consistently with this approximation, we first neglect the $(1/b)$ terms in the last factors of Eqs.(5a) and (6a), in which limit these expressions coincide. This approximation is mathematically equivalent to an approach sometimes referred to as "naive non–abelianization"[8]. The effect of retaining the specific $(1/b)$ contributions in Eq.(5a), and in Eq.(6a) for special $\xi_Q$ values, will be discussed later.

The sum of contributions involving $n = 1, 2, ...., \infty$ bubbles is a geometric series that renormalizes $(g_s^0)^2$ according to

$$(g_s^0)^2 \longrightarrow \frac{(g_s^0)^2}{1-\widehat{\Pi}(k^2)} \ . \tag{7}$$

In order to express this amplitude in terms of renormalized parameters, we write

$$(g_s^0)^2 = 4\pi \left(\frac{\mu^2 e^\gamma}{4\pi}\right)^\epsilon \frac{\alpha_s(\mu)}{Z_3} \ , \tag{8}$$

so that the r.h.s. of Eq.(7) becomes

$$\frac{(4\pi\alpha_s(\mu)/Z_3)\,(\mu^2 e^\gamma/4\pi)^\epsilon}{\left[1 - 3\,(b\alpha_s(\mu)/2\pi Z_3)\,(-\mu^2 e^\gamma/k^2)^\epsilon\,\Gamma(\epsilon)\mathrm{B}(2-\epsilon, 2-\epsilon)\right]} \ . \tag{9}$$

In the $\overline{\mathrm{MS}}$ renormalization, $Z_3$ is adjusted to cancell the $1/\epsilon$ singularity in the second term of the denominator. Thus, for the diagrams under consideration,

$$Z_3 = 1 + \frac{ba(\mu)}{4\epsilon} \ . \tag{10}$$

Setting $-k^2 = \kappa^2$, we note parenthetically that in the $\epsilon \to 0$ limit Eq.(7) reduces to $\alpha_s(e^{-5/6}\kappa)$, the V-scheme running coupling [14].

We now turn our attention to the mass renormalization

$$M - m_0 = \Sigma(\not{p} = M) \ , \tag{11}$$

where $m_0$ is the bare mass and $\Sigma$ the quark self–energy. At the one–loop level we have

$$\Sigma(\not p = M) = -i\frac{4}{3}\int \frac{d^n k}{(2\pi)^n}\frac{(g_s^0)^2(2M - \not k(n-2))}{k^2(k^2+2p\cdot k)}\Bigg|_{\not p = M} , \tag{12}$$

where, following the derivation of Eqs.(1,2) [12], we have neglected width effects. In order to absorb the vacuum polarization contributions, the next step is to replace $(g_s^0)^2$ in Eq.(12) by the expression of Eq.(9), expand the geometric series and carry out the $n$–dimensional integration. The result is

$$\Sigma(\not p = M) = \frac{4M}{b}\sum_{n=1}^{\infty}\left(\frac{ba(\mu)}{4\epsilon Z_3}\right)^n \frac{f(\epsilon, n\epsilon)}{n} , \tag{13}$$

where

$$\begin{aligned} f(\epsilon, y) \equiv \; & \left(\tfrac{\mu^2}{M^2}\right)^y \left[6e^{\gamma\epsilon}\Gamma(1+\epsilon)\mathrm{B}(2-\epsilon, 2-\epsilon)\right]^{\frac{y}{\epsilon}-1} \\ & \times\, 2e^{\gamma\epsilon}\Gamma(1+y)\Gamma(1-2y)(1-y)(1-\tfrac{2\epsilon}{3})/\Gamma(3-\epsilon-y) . \end{aligned} \tag{14}$$

Recalling Eq.(10), we now expand $(1/Z_3)^n$ in powers of $\chi(\mu) \equiv ba(\mu)/4\epsilon$ and re–arrange the double summation to read

$$\Sigma(\not p = M) = \frac{4M}{b}\sum_{n=1}^{\infty}(\chi(\mu))^n (n-1)! \sum_{l=1}^{n}\frac{(-1)^{n-l}}{l!(n-l)!} f(\epsilon, l\epsilon) . \tag{15}$$

Next we substitute $M \to m_0$ in the overall cofactor, as the difference involves terms of $O(b^l \alpha_s^n)$ $(l \le n-2)$ which are beyond our approximation. Inserting the result in Eq.(11), performing the $\overline{\mathrm{MS}}$ –renormalization

$$m_0 = \hat{m}(\mu) Z_{(m)} , \tag{16}$$

where

$$Z_{(m)} = 1 + a(\mu)\frac{Z_1}{\epsilon} + a^2(\mu)\left(\frac{Z_{22}}{\epsilon^2} + \frac{Z_{21}}{\epsilon}\right) + \dots , \tag{17}$$

and dividing by $\hat{m}(\mu)$ we find

$$\frac{M}{\hat{m}(\mu)} = Z_{(m)}\left[1 + \frac{\Sigma(\not p = M)}{M}\right] . \tag{18}$$

The counterterms present in $Z_{(m)}$ are important to cancel $\epsilon$ singularities but it is easy to see that they only contribute finite terms of $O(b^l \alpha_s^n)$ $(l \leq n-2)$. For example, $Z_1 = O(1)$ and contributes finite terms of $O(a^2)$ but not $O(ba^2)$; $Z_{22}$ and $Z_{21}$ contain terms of $O(b)$ and contribute finite terms of $O(ba^3)$ but not $O(b^2a^3)$. Thus, in the finite terms of $O(b^{n-1}a^n)$ the $\epsilon^n$ contained in $\chi^n(\mu)$ are cancelled by $\epsilon$ factors from $f(\epsilon, l\epsilon)$. The $l$ summation in Eq.(15) can be carried out by expanding

$$f(\epsilon, l\epsilon) = \sum_{p=0}^{n} c_p(\epsilon)(l\epsilon)^p + \mathcal{O}(\epsilon^{n+1}) \ , \tag{19}$$

interchanging the $l$ and $p$ summations so that

$$\sum_{l=1}^{n} \frac{(-1)^{n-l}}{l!(n-l)!} f(\epsilon, l\epsilon) = \sum_{p=0}^{n} c_p(\epsilon)\epsilon^p \sum_{l=1}^{n} \frac{(-1)^{n-l}}{l!(n-l)!} \ l^p + \mathcal{O}(\epsilon^{n+1}) \ , \tag{20}$$

and using the identity

$$\sum_{l=1}^{n} \frac{(-1)^{n-l}}{l!(n-l)!} \ l^p = \frac{(-1)^{n+1}}{n!}\delta_{p0} + \delta_{pn} \qquad\qquad (p \leq n) \ . \tag{21}$$

Thus

$$\sum_{l=1}^{n} \frac{(-1)^{n-l}}{l!(n-l)!} f(\epsilon, l\epsilon) = \frac{(-1)^{n+1}}{n!} c_0(\epsilon) + c_n(\epsilon)\epsilon^n + \mathcal{O}(\epsilon^{n+1}) \ . \tag{22}$$

The terms of $\mathcal{O}(\epsilon^{n+1})$ give vanishing contributions to Eq.(15) as $\epsilon \to 0$. Eq.(21) can be proved by applying the differential operator $\left(x\frac{d}{dx}\right)^p$ to both sides of the binomial expansion

$$(1-x)^n = \sum_{l=0}^{n} \binom{n}{l} (-1)^l x^l \ , \tag{23}$$

and then setting $x = 1$. Inserting Eq.(22) into Eqs.(15,18), we have

$$\frac{M}{\hat{m}(\mu)} = Z_{(m)} \left\{ 1 + \frac{4}{b} \sum_{n=1}^{\infty} \left( \frac{a(\mu)b}{4} \right)^n (n-1)! \left[ c_n(\epsilon) + \frac{(-1)^{n+1}}{n!} \frac{c_0(\epsilon)}{\epsilon^n} \right] \right\} \ . \tag{24}$$

From Eqs.(14,19) it follows that

$$c_0(\epsilon) = f(\epsilon, 0) = \left( \frac{1 - 2\epsilon/3}{1-\epsilon} \right)^2 \left( \frac{1-2\epsilon}{1-\epsilon/2} \right) \frac{\Gamma(1-2\epsilon)}{\Gamma(1+\epsilon)\Gamma^3(1-\epsilon)} \ . \tag{25}$$

Expanding

$$f(\epsilon,0)=\sum_{r=0} f_r \epsilon^r \ , \tag{26}$$

and neglecting terms of $\mathcal{O}(b^l {\alpha_s}^n)$ $(l \leq n-2)$, only $f_n$ and $c_n(0)$ contribute to Eq.(24) as $\epsilon \to 0$. Thus

$$\frac{M}{\hat{m}(\mu)} = 1+\frac{4}{b}\sum_{n=1}\left(\frac{a(\mu)b}{4}\right)^n (n-1)!\left[c_n(0)+\frac{(-1)^{n+1}}{n!}\ f_n\right] \ , \tag{27}$$

where the $\epsilon$ singularities have explicitly cancelled. The $c_n(0)$ can be obtained by using again Eqs.(14,19) :

$$f(0,y)=\left(\frac{\mu^2}{M^2}e^{5/3}\right)^y \frac{\Gamma(1+y)\Gamma(1-2y)}{(1-y/2)\Gamma(1-y)}=\sum_{n=0} c_n(0)y^n \ . \tag{28}$$

We note that the $c_n(0)$ depend on $\mu^2/M^2$. It is interesting to observe that, using Eq.(28), the Borel transform of the first series in Eq.(29), involving $c_n(0)$, can be expressed in closed analytic form. In fact, it is given by

$$\frac{1}{\pi}\sum_{n=0}^{\infty}\left(\frac{bt}{4\pi}\right)^n c_{n+1}(0)=\frac{1}{\pi}\frac{[f(0,u)-1]}{u}=\frac{1}{\pi}\left[\left(\frac{\mu^2}{M^2}e^{5/3}\right)^u \frac{2(1-u)\Gamma(u)\Gamma(1-2u)}{\Gamma(3-u)}-\frac{1}{u}\right] \ ,$$

where $u=bt/4\pi$ and $t$ is the Borel parameter. This expression exhibits the characteristic infrared renormalon pole at $u=1/2$, and has been already given by Beneke and Braun in the leading $1/n_f$ approximation (cf. Eq.(3.13) of Ref. [7]). We have not found, however, a simple closed form for the Borel transform of the second series, involving $f_n$. For our purposes, which is to evaluate the coefficients of $a^n$, Eq.(27) suffices.

In summary, the terms of $\mathcal{O}(b^{n-1}{\alpha_s}^n)$ are given by Eq.(27). The coefficients $f_n$ and $c_n(0)$ are obtained form Eqs.(25,26,28). Setting $\mu=M$, Eq.(27) can be expressed as

$$\frac{M}{\hat{m}(M)}=1+\frac{4a(M)}{3}\sum_{n=0}^{\infty}\left(\frac{ba(M)}{2}\right)^n n!A_n \ , \tag{29}$$

where

$$A_n=\frac{3}{2^{n+2}}\left[c_{n+1}(0)+\frac{(-1)^n}{(n+1)!}\ f_{n+1}\right] \ . \tag{30}$$

The coefficients $A_n$ are given in Table 1 up to $n = 10$. We note that the $A_n$ are remarkably constant for $n \geq 1$ and converge to a value $\approx 2.301$ very close to $e^{5/6}$, a factor associated with the residue of the infrared renormalon pole. In particular, $A_1 = \pi^2/8 + 71/64$. The second order term is $2bA_1a^2/3 = 1.56205ba^2$.Recalling that $b$ contains $-2n_f/3$, we see that the coefficient of $n_f$ in the $a^2$ term is –1.0414, in agreement with Eq.(2). This is a welcome check, as our calculation should reproduce this factor exactly. For the top, bottom and charm cases our approximate evaluation of the $a^2$ term leads to coefficients 11.98, 13.02, and 14.06, respectively, to be compared with 10.90, 11.94, and 12.99 from Eq.(2) when the small $\Delta$ corrections are neglected. We see that the $n = 0, 1$ terms in Eq.(29) approximate very well the exact results, thus providing a simple physical interpretation of these important but technically complex corrections.

At this stage we return to Eq.(5a), and examine the effect of retaining the specific $1/b$ term contained in the last factor. It is easy to see that the only changes are an additional factor $[1 + 4\epsilon/b(1-\epsilon)]^{-1}$ in the r.h.s. of Eq.(25) and the replacement $e^{5/3} \rightarrow e^{5/3+4/b}$ in Eq.(28). This leads to an expansion of the same form as Eq.(29) with $A_n \rightarrow P_n$, where the first ten coefficients are given in Table 2 in the top quark case ($b = 23/3$). We see that the $P_n$ are asymptotically larger than the $A_n$ by $e^{2/b} \approx 1.30$, with smaller differences for low $n$ values. In particular, the $\mathcal{O}(a^2(M))$ coefficient in $M/\hat{m}(M)$ is now $1.56205\, b + 9/4$. It contains again the exact $n_f$ term and, in the top–quark case, amounts to 14.23, which is 19% larger that the result 11.98 from Eq.(29) and 30% larger than the value 10.90 from the exact calculation of Eqs.(1,2). Thus, the $P_n$ expansion provides a rough aprroximation to Eqs.(1,2) through $\mathcal{O}({\alpha_s}^2)$, but it is not nearly as precise as Eq.(29). It is also interesting to inquire whether the aprroximation of neglecting the $1/b$ terms in Eqs.(5a) or (6a) corresponds exactly to a specific choice of gauge in the BFG formulation. We note that the gauges $\xi_Q = -3 \pm 4\sqrt{2/3}$ reduce the expression between curly brackets in Eq.(6a) to $1 + (8/3b)(\epsilon^2/(1-\epsilon))$. As the cofactor of this expression in Eq.(5a) has only a $1/\epsilon$ singularity, in four dimensions these particular choices of $\xi_Q$ correspond, in fact, to that approximation. However, the presence of the residual $\epsilon^2$ contributions shows that in $n-$dimensions there exists no BFG that exactly reproduces the "naive non–abelianization" prescription. The retention of this $\epsilon^2-$term does not affect Eq.(28), but it introduces an additional factor $[1 + 8\epsilon^2/3b(1-\epsilon)]^{-1}$ in Eq.(25). This alters the $f_r$ parameters and leads to an expansion of the same form as Eq.(29), with slightly modified coefficients. In particular the $\mathcal{O}(a^2(M))$ coefficient in $M/\hat{m}(M)$ becomes $1.56205\ b + 1/3$. On the other hand, as the $c_n(0)$ are not affected, the asymptotic behaviour is the same as in the $A_n$ expansion.

As a further application of Eq.(6a), we note that any choice of gauge with $\xi_Q \geq 1$ will increase the coefficient of the $\epsilon$ term in the expression between curly brackets, relative to the PT result, and therefore lead to asymptotically larger coefficients. In the following we consider three interesting cases in which the opposite occurs, namely $\xi_Q = 0$ (which may be referred to as the transverse or Landau BFG), $\xi_Q = -3$ and a specific gauge $\xi_Q^*$ to be defined later. The $\xi_Q = -3$ gauge corresponds to the minimum of the $\epsilon$ cofactor in Eq.(6a) and, therefore, to expansion parameters with the smallest asymptotic behaviour. By a curious coincidence, in four dimensions the $\xi_Q = -3$ BFG self–energy also coincides with the one–loop correction to the interaction potential between two infinitely heavy quarks (see discussion after Eq.(31)). The $\xi_Q^*$ gauges are defined so that the $\mathcal{O}(a^2(M))$ contribution in Eq.(29) coincides with the exact value $K$ (Eq.(2)) when the small $\Delta-$terms are neglected. In other words, the complete $\mathcal{O}(a^2(M))$ correctios to $M/\hat{m}(M)$ is identified as a self–energy contribution. This can be readily implemented by noting that the contribution from Eq.(6a) to the $a^2(M)$ coefficient in Eq.(29) is $1.56205\ b + (9 - 23c/3)/4$, where $c = 3(1-\xi_Q)(7+\xi_Q)/16$, while Eq.(2) with high precision can be rewritten as $1.56205b - 1.07249$. Thus, $c = 1.73347$, which corresponds to $\xi_Q^* \approx -5.599$ or $-0.401$. We also note that in the general $\xi_Q$ gauge, the asymptotic behaviour of the corresponding coefficients in Eq.(29) is $e^{5/6+2(1-c)/b}$.

For $\xi_Q = 0$, the factor between curly brackets in Eq.(6a) reduces to the expression $1 - [\epsilon/b(1-\epsilon)][5/4 - 7\epsilon/2]$. This introduces a modification $e^{5/3} \to e^{5/3-5/4b}$ in Eq.(28) and an additional factor $[1 - (\epsilon/b(1-\epsilon))(5/4 - 7\epsilon/2)]^{-1}$ in Eq.(25). The corresponding $L_n$ coefficients are shown in Table 2 for the top quark case. We see that the asymptotic behaviour is reduced by a factor $e^{-5/8b} = 0.922$ relative to that in Eq.(29). On the other hand, the $\mathcal{O}(a^2(M))$ coefficient in $M/\hat{m}(M)$ becomes $1.56205b - 0.26563$, which is very close to the result in Eq.(29).

For general $\xi_Q$, we have the modification $e^{5/3} \to e^{5/3+4(1-c)/b}$ in Eq.(28) and an additional factor $\{1 + [4\epsilon/b(1-\epsilon)][1 - c(1 - 2\epsilon/3)]\}^{-1}$ in Eq.(25). As mentioned before, in the $\xi_Q^*$ case $c = 1.73347$ and we obtain the $B_n^*$ coefficients shown in Table 2 for the top quark. The asymptotic behaviour is now reduced by $e^{-1.467/b} = 0.826$ relative to that in Eq.(29). The $\xi_Q^*$ prescription has the interesting feature of incorporating into Eq.(29) the complete $\mathcal{O}(a^2(M))$ contribution of Eq.(2), which is manifestly gauge invariant.

For $\xi_Q = -3$, the expression between curly brackets in Eq.(6a) simplifies to $1 - 8\epsilon/b$. Recalling Eq.(8), neglecting the higher–order $1/Z_3$ correction and taking the $\epsilon \to 0$ limit, we have

$$\Pi(k^2, -3)|_{BFG}^{\overline{\rm MS}} = \frac{\alpha_s(\mu)}{4\pi} b \left[ \ln(-\frac{\mu^2}{k^2}) + \frac{5}{3} - \frac{8}{b} \right] , \qquad (31)$$

where the superscript $\overline{\rm MS}$ means that the $\overline{\rm MS}$ renormalization has been implemented. Eq.(31) coincides with the one–loop correction to the interaction potential between two infinitely massive quarks [21], and thus it is amenable to direct physical interpretation. The $1 - 8\epsilon/b$ correction induces a modification $e^{5/3} \to e^{5/3-8/b}$ in Eq.(28) and an additional factor $(1 - 8\epsilon/b)^{-1}$ in Eq.(25). The corresponding $V_n$ coefficients are shown in Table 5. Asymptotically, $V_n \sim e^{-4/b} A_n = 0.59 A_n$ in the top quark case, a significant difference. The $\mathcal{O}(a^2(M))$ coefficient in $M/\hat{m}(M)$ becomes $1.56205b - 7/2$, which in the top quark case is 22% smaller than the exact result of Eq.(2).

In summary, if the basic one–loop vacuum polarization diagram is defined on the basis of Eqs.(5a) or (6a) with the neglect of the $1/b$ terms, the contributions of $\mathcal{O}(b^{n-1}{\alpha_s}^n)$ to $M/\hat{m}(M)$ are given by the $A_n$ expansion of Eq.(29), which corresponds to the so called "naive non–abelianization" prescription. If the $1/b$ terms in Eq.(5a) are retained, in accordance with a definition of the basic one–loop vacuum–polarization function based on the complete PT self–energy , we have the alternative expansion involving the $P_n$ coefficients . They are (19–30)% larger than their $A_n$ counterparts. If instead, the $1/b$ terms in Eq.(6a) are retained, corresponding to the BFG framework, the answer depends on the choice of $\xi_Q$. We have discussed five particular cases : i) $\xi_Q = 1$ which also leads to the $P_n$ expansion. ii) $\xi_Q = -3 \pm 4\sqrt{2/3}$, which is very close to the $A_n$ case. iii) $\xi_Q = 0$ (Landau or transverse BFG) with coefficients $L_n$ similar to the $A_n$, but asymptotically reduced by $e^{-5/8b} = 0.922$. v) $\xi_Q^*$ which incorporates the exact result of $\mathcal{O}(a^2(M))$. iv) $\xi_Q = -3$ which leads to expansion coefficients $V_n$ with the smallest asymptotic limit. In all cases we have the important constraint that the $a^2(M)$ coefficients should be close to the exact answer of Eq.(2). According to this criterion, the $\xi_Q^*$ BFG calculation is singled out, as it reproduces Eq.(2) exactly. However, the $A_n$ and $L_n$ expansions also fare very well, with $\mathcal{O}(a^2(M_t))$ coefficients that are only 9.8% and 7.4% larger, respectively, than the exact result. On the other hand, the $P_n$ and $V_n$ expansions are not so accurate, with $\mathcal{O}(a^2(M_t))$ coefficients 30% higher, and 22% lower, respectively. As there are other contributions of $\mathcal{O}(b^l{\alpha_s}^n)$ $(l \leq n-2)$ which we have neglected, it may be argued that the $A_n$ expansion represents a more self–consistent approach than its $P_n$, $L_n$ and $V_n$ counterparts. On the other hand, the $\xi_Q^*$ BFG prescription has the advantage of exactly incorporating the most important subleading contributions, namely those of $\mathcal{O}(a^2(M))$. It is clear that the $\xi_Q^*$ procedure can be generalized to other QCD calculations where the $a^2(M)$ coefficients are known exactly.

After completing our calculation, we have learned that very recently [8] Beneke and Braun have also considered the detailed resummation of fermion bubble chains ( characterized as the leading corrections in the $1/n_f$ expansion ) and proposed to include the gluonic contribution by the heuristic procedure $n_f \to -3b/2$, which is refered to as "naive non–abelianization". In the $M/\hat{m}(M)$ calculation, this approach should be equivalent to the $A_n$ expansion of Eq.(29). Their result is expressed in the form

$$\frac{M}{\hat{m}(M)} = 1 + \frac{a(M)}{3} \sum_{n=0}^{\infty} r_n \left( \frac{ba(M)}{4} \right)^n ,$$

where the large $r_n$ coefficients are evaluated by ingenious Borel transform techniques and given numerically up to $n = 8$. We have verified that the relation $r_n = 2^{n+2} n! A_n$, necessary for the consistency of the results of Ref.[8] with Eq.(29), holds numerically with high precision up to $n = 8$.

Assuming that the terms $n \geq 2$ also give dominant contributions, we employ Eq.(29) to estimate the higher order coefficients and evaluate $M/\hat{m}(M)$. For the top quark case, according to Eq.(29) the $O(a^3)$ and $O(a^4)$ coefficients are 86 and 1031, which are close to the values $\approx$ 104 and 1041 estimated by optimization methods [1,2]. Their sizeable magnitude arises because of the factorial growth and the fact that the expansion parameter involves the rather large factor $b/2$. As all the terms are of the same sign, Eq.(29) with asymptotically constant $A_n$ is not Borel–summable. Therefore, the sum is carried out to the optimal term of smallest magnitude, which at the same time is regarded as an irreducible theoretical error associated with the presumed asymptotic nature of the perturbative QCD expansion [5–11]. Although the $O(a^2)$ term in Eq.(29) is a very good approximation, in order to obtain a more accurate answer we replace it, in the top–quark case, by $10.90a^2$, the $O(a^2)$ contribution from the exact Eq.(2) when the small $\Delta$ corrections are neglected. As an illustration, we consider $M_t = 175$ GeV. Using a three–loop $\beta$ function with 5 active flavours normalized to $\alpha_s(m_Z) = 0.118$, we have $\alpha_s(175\text{GeV}) = 0.1074$. In this example, the smallest term in Eq.(29)

occurs at $n = 7$ and we obtain $M_t/\hat{m}_t(M_t)$= 1.06518 if this term is included, and 1.06482 if the series is terminated at $n = 6$. This is to be compared with the values 1.06470, 1.06470, and 1.06463 derived using the three optimization methods [1,2], and 1.05835 from Eq.(1). Thus, the QCD correction in the present calculation (0.06482–0.06518) is very close to the results found by optimization methods, and differs from the results from Eq.(1) by (11–12)%. Table 3 extends this comparison to the range 130 GeV $\leq m_t \leq$ 220GeV. The second column gives the value from Eq.(1), the third column that from the BLM expression (Eq.(3)), the fourth and fifth columns give the results of incorporating the higher order terms in Eq.(29) up to $n = 6$ and $n = 7$ (previous to smallest and smallest terms, respectively). The smallest term for $M_t$ = 175GeV is $3.52\times10^{-4}$, so that our estimate of the irreducible theoretical error in $M_t$ is $3.52 \times 10^{-4} \times 175\text{GeV}/1.065 \approx 58\text{MeV}$. If we use Stirling's approximation and the one–loop relation $2/ba(M) = \ln(M/\Lambda_{\rm QCD})$, our irreducible error estimate can be written as

$$\delta M \approx \frac{8}{3}\left(\frac{\alpha_s(M)}{b}\right)^{1/2} e^{5/6}\Lambda_{\rm QCD} \tag{32}$$

In this case $\Lambda_{\rm QCD}\approx$ 85 MeV (from the one–loop relation) and from (32) one has $\delta M_t \approx 62$ MeV. An alternative procedure to estimate $\delta M$ is based on the ambiguity of the Borel transform. For example, using Eq.(34) of Ref.[8], one obtains an expression similar to Eq.(32) with $(\alpha_s(M)/b)^{1/2} \to 1/b$. In the top quark case this increases the uncertainty by a factor $(\alpha_s(M)\ b)^{-1/2} \approx 1.10$ so that $\delta M \approx 68$ MeV. Fortunately, such uncertainties are phenomenologically negligible. Because complete higher order results are not available, the actual error in current calculations is, of course, likely to be significantly higher. In Ref.[2] the last term in Eq.(3) was used as an estimate of the magnitude of the theoretical uncertainty and this gives $\pm2.74 \times 10^{-3}$. It is worth noting that an analogous use of the $P_n$ and $V_n$ expansions, i.e. the two extreme cases we have considered, lead to values of $M_t/\hat{m}_t(M_t)$ that differ from the BLM result (Eq.(3)) by $(2 - 2.5) \times 10^{-3}$. On the other hand, the $L_n$ and $B^*_n$ expansions differ from the BLM result by $\lesssim 1 \times 10^{-3}$. We also emphasize that the optimization methods applied directly to the $O(\alpha_s^2)$ expansion (Eq.(1)) fail to uncover the factorial growth displayed in Eq.(29). For instance, if $a(\mu_t^*)$ in Eq.(3) is expanded in terms of $a(M_t)$, it obviously generates a geometric series. The numerical closeness of the two calculations is therefore very interesting.

We briefly illustrate the application of Eq.(1) and Eq.(29) to the bottom quark case using $M_b = 4.72$ GeV [13], in which case $\alpha_s(4.72\text{GeV}) = 0.2162$. We again improve the accuracy by replacing the $a^2$ term in Eq.(29) by $12.5a^2$, the result obtained from the exact Eq.(2) including the $\Delta$ corrections. In this case, the smallest term is $n = 3$, and we have $M_b/\hat{m}_b(M_b) = 1.214$ if it is included, and 1.184 if the sum is stopped at $n = 2$. This is to be compared with 1.151 from Eq.(1). Such differences are not too surprising. We note that the expansion from Eq.(1) is 1 + 0.0918 + 0.0592. Thus, although Eq.(1) is routinely applied to the $b$ and $c$ quarks [4], its convergence properties suggests a large theoretical uncertainty in those cases. As the value of $\hat{m}_b(M_b)$ is frequently derived from $M_b$ by using Eq.(1) [4], one obtains in this case $\hat{m}_b(M_b)$ = 4.72 GeV/1.151 = 4.10 GeV, while application of the present $n = 2$ calculation leads to $\hat{m}_b(M_b)$ = 4.72 GeV/1.184 = 3.99 GeV, a difference of 114 MeV or $\approx 3\%$. The smallest term in the series amounts in this case to $2.97\times10^{-2}$, so that our estimate of the irreducible theoretical uncertainty in $M_b$ is $2.97\times10^{-2}\times3.99$ GeV $\approx$ 119 MeV, which is twice as large as that given in Ref.[6]. This difference can be roughly traced to the factor $e^{5/6}$ in Eq.(32). A more detailed treatment of $M/\hat{m}(M)$ for the $b$ and $c$ cases has recently been proposed by Beneke and Braun [8]. In the $b$ case their central value is close to the average of the two values we have given, while their irreducible error is about 25% smaller.

## 2 QCD corrections to $(\Delta\rho)_f$

The study of the QCD corrections to $(\Delta\rho)_f$, the fermionic component of $\Delta\rho$, is a subject of considerable interest, as this amplitude contains the leading asymptotic contributions, for large $M_t$, of the basic radiative corrections $\Delta r$,$\Delta\hat{r}$, and $\Delta\hat{\rho}$ [22,23]. We also recall that $\rho_f \equiv [1-(\Delta\rho)_f]^{-1}$ is frequently separated, as an overall renormalization factor, when neutral current amplitudes are expressed in terms of $G_\mu$. In this section we apply the previous results to the study of these corrections. We also update the discussion of Refs.[1,2], including an estimate of the theoretical error, to take into account very recent modifications in the $\mathcal{O}(\alpha\alpha_s^2)$ calculations.

Neglecting higher–order electroweak effects $\sim \mathcal{O}(\alpha/\pi s^2)^2(M_t^2/m_W^2)$, but retaining QCD corrections, $(\Delta\rho)_f$ can be written as

$$(\Delta\rho)_f = \frac{3G_\mu M_t^2}{8\sqrt{2}\pi^2}\left[1+\delta_{\rm QCD}\right] \ , \tag{33}$$

where the first factor is the one–loop result and $\delta_{\rm QCD}$ represents the relevant QCD correction. Alternatively, calling $\mu_t$ the solution of $\hat{m}(\mu) = \mu$, where $\hat{m}_t(\mu)$ is the $\overline{\rm MS}$ –running mass, Eq.(33) can be expressed in the equivalent forms

$$(\Delta\rho)_f = \frac{3G_\mu \mu_t^2}{8\sqrt{2}\pi^2}\left[1+\delta_{\rm QCD}^{\overline{MS}}\right] \ , \tag{34}$$

and

$$(\Delta\rho)_f = \frac{3G_\mu \hat{m}_t^2(M_t)}{8\sqrt{2}\pi^2}\left[1+\Delta_{\rm QCD}\right] \ . \tag{35}$$

Eq.(34) can be interpreted as a pure–$\overline{\rm MS}$ expression, while Eq.(35) is very useful because it can be used in conjunction with the results of Section 1.

Two recent evaluations of the complete three–loop corrections of $\mathcal{O}(\alpha\alpha_s^2)$ have been given by Avdeev, Fleischer, Mikhailov, and Tarasov [24], and by Chetyrkin, Kühn, and Steinhauser [25], in the limit $M_b = 0$. After an initial discrepancy the two results now agree and, to good accuracy, are given by

$$\delta_{\rm QCD} = -2.8599\ a(M_t) - (23.525 - 1.7862\ n_f)a^2(M_t) \ , \tag{36}$$

where the $\mathcal{O}(a(M_t))$ term is the Djouadi–Verzegnassi result, and $n_f = 5$ is the number of massless flavours. Numerically, the second order coefficient is $-14.594$. Using Eq.(1) in the

$\Delta = 0$ approximation and the relation $\mu_t = [1 + (8/3)a^2(M_t) + ...]\hat{m}(M_t)$, Eq.(36) implies

$$\delta^{\overline{MS}}_{\rm QCD} = -0.19325\ a(M_t) - 3.970\ a^2(M_t)\ , \tag{37}$$

$$\Delta_{\rm QCD} = -0.19325\ a(M_t) + 1.364\ a^2(M_t)\ . \tag{38}$$

As pointed out in Ref.[25], the second order coefficient in Eq.(37) is almost entirely due to the opening of a new channel in $\mathcal{O}(\alpha\alpha_s^2)$, namely the double triangle graph. It is interesting to compare the new results of Eqs.(36–38) with a very simple estimate made before the complete three–loop calculations were carried out [26]. In that analysis the correction $\Delta_{\rm QCD}$ was written as

$$\Delta_{\rm QCD} = -0.19325\ a(M_t) + C\ a^2(M_t)\ , \tag{39}$$

and the constant $C$ estimated to be $C = 0 \pm 6$ on the basis of convergence assumptions (optimization arguments carried out in Ref.[26] also led to $C \approx +3$, but the more conservative estimate $0 \pm 6$ was employed in the analysis). In turn, $C = 0 \pm 6$ implies a second–order coefficient $-15.958 \pm 6$ in Eq.(36). Thus, we see that the central values in the estimate of Ref.[26], namely C=0 in Eq.(39) and $-15.958$ in Eq.(36), are amusingly close to the new revised second–order coefficients in the exact calculation .

Applying directly the optimization methods to Eq.(36), we find

$$\delta_{\rm QCD} = -2.8599\ a(0.1536M_t) + 5.947\ a^2(0.1536M_t) \qquad \text{(BLM)}\ , \tag{40}$$

$$\delta_{\rm QCD} = -2.8599\ a(0.2241M_t) + 1.803\ a^2(0.2241M_t) \qquad \text{(PMS)}\ , \tag{41}$$

$$\delta_{\rm QCD} = -2.8599\ a(0.2642M_t) \qquad \text{(FAC)}\ . \tag{42}$$

Eqs.(40–42) can be obtained, for example, by using the explicit expressions given in Eqs.(16–18) of Ref.[2]. As an illustration, we consider the case $M_t = 175$ GeV. Employing a three–loop $\beta$ function with five active flavours ( we neglect the very small three–loop discontinuity at $\mu = M_t$) and normalizing once more $\alpha_s(\mu)$ such that $\alpha_s(M_Z) = 0.118$ for $M_Z = 91.19$ GeV,

Eqs.(40–42) give, for $M_t = 175$ GeV, $\delta_{\rm QCD} = -0.1192$, $-0.1198$, and $-0.1197$, respectively. This is to be compared with $-0.1149$ from Eq.(36). It is interesting to note that: i) The BLM approach works considerably better with the recently modified version of $\delta_{\rm QCD}$, in the sense that the residual $\mathcal{O}(a^2)$ term is not as large as before. ii) The three optimization methods give now close answers. iii) The difference between the results from the optimization methods and Eq.(36) is $\approx 4.9 \times 10^{-3}$, which is roughly as large as the effect of the recent modifications, namely $4.04\ a^2(M_t) = 4.7 \times 10^{-3}$. It should be stressed that the $4.9 \times 10^{-3}$ variation is due to different ways of accounting for the contributions of $\mathcal{O}(a^3)$ and higher.

The presence of large and increasing coefficients, the observations in iii), and the large vacuum–polarization contributions to $M_t/\hat{m}(M_t)$ we have encountered in Section 1, strongly suggest that the corrections of $\mathcal{O}(a^3)$ and higher to Eq.(36) are significant. It is therefore a good idea to examine alternative evaluations of $\delta_{\rm QCD}$ that involve second order coefficients of $\mathcal{O}(1)$, rather than $\mathcal{O}(10)$. Such strategy was explained in Refs.[1,2], and is updated in the following to take into account the very recent modifications in the $\mathcal{O}(\alpha\alpha_s^2)$ corrections and the results of Section 1.

Combining Eqs.(33) and (35) we have

$$1 + \delta_{\rm QCD} = \left(\frac{\hat{m}(M_t)}{M_t}\right)^2 [1 + \Delta_{\rm QCD}]\ . \tag{43}$$

The correction $\Delta_{\rm QCD}$ is very small. However, in order to incorporate large vacuum–polarization effects that are induced in $\mathcal{O}(a^3)$, we replace Eq.(38) by the more detailed expression

$$\Delta_{\rm QCD} = -0.19325\ a(M_t) - 3.970\ a^2(M_t) + \frac{16}{3}a^2(0.252\ M_t) - 9.97\ a^3(0.252\ M_t)\ . \tag{44}$$

This is obtained by combining Eq.(37) with the BLM optimized expansion

$$\frac{\mu_t}{\hat{m}_t(M_t)} = 1 + \frac{8}{3}a^2(0.252\ M_t) - 4.47a^3(0.252\ M_t)\ , \tag{45}$$

derived in Ref.[1]. The large vacuum–polarization effects of $\mathcal{O}(a^3)$ incorporated in Eq.(45) are due to the presence of the pole mass in the argument of $\hat{m}$. As Eq.(45) is exactly known through $\mathcal{O}(a^3)$ [1], this is a very instructive illustration of the fact that the occurence of the pole mass $M$ in theoretical expressions is very likely to engender large vacuum–polarization effects, as well as large coefficients in higher orders, if $\alpha_s(M)$ is employed as an expansion parameter. The correction $\Delta_{\rm QCD}$, evaluated on the basis of Eq.(44), is given as a function of $M_t$, in Table 4. We see that $\Delta_{\rm QCD} \approx -(2.3-2.6)\times 10^{-3}$ is very small. This is close in magnitude to the values reported before [1], but of opposite sign. The difference is due to the modification of the second–order coefficients in the exact calculation .

Table 5 compares several evaluations of $\delta_{\rm QCD}$, as a function of $M_t$. The second column gives the values from Eq.(36) while the remaining ones report the results from using Eq.(43), Table 4 for $\Delta_{\rm QCD}$, and the evaluations of $M_t/\hat{m}_t(M_t)$ given in Section 1. Specifically, the third, fourth, and fifth column employ the $M_t/\hat{m}_t(M_t)$ values from the BLM approach (Eq.(3)), and the $n=6$ and $n=7$ summations explained in Section 1, respectively (see Table 3). It is clear that the last three columns, and especially the third and fourth, are quite close. On the other hand, they are larger than the first by roughly (4–6)% depending on $M_t$ and the particular entries being compared. This is not too surprising because the last three columns incorporate relatively large vacuum–polarization contributions of $\mathcal{O}(a^3...)$ to $M_t/\hat{m}_t(M_t)$, not included in the first one. Because of the closeness of the last three columns, there is no point in drawing distinctions among them and, for definiteness, we adopt the entries in the third one, based on Eq.(43) and the BLM optimization of Eq.(3), as our central values. Following Ref.[2], we estimate the theoretical error due to unknown higher order corrections by the magnitude of the last retained terms. We emphasize that, although the three optimizations of $M_t/\hat{m}_t(M_t)$give nearly the same numerical answer, in order to analyze the error we employ the BLM approach, which leads to the most conservative estimate. From Eq.(3) we have

$$\delta\left(\frac{\hat{m}(M_t)}{M_t}\right)^2 \approx \pm\frac{2\times 1.073 a^2(\mu^*)}{(M_t/\hat{m}(M_t))^3} \approx \pm 1.78\ a^2(\mu^*)\ . \tag{46}$$

From Eq.(38)

$$\delta(\Delta_{\rm QCD}) \approx \pm 10\ a^3(0.252\ M_t)\ . \tag{47}$$

Inserting these uncertainties in Eq.(43) and adding them linearly we obtain

$$\delta(\delta_{\rm QCD}) = \pm(4.8,\ 4.9,\ 5.2,\ 5.9) \times 10^{-3}\ , \tag{48}$$

for $M_t = 220,\ 200,\ 175,\ 130$ GeV, respectively. These error estimates coincide with the values reported in Ref.[2].

In analogy with the discussion in Refs.[1,2], the values for $\delta_{\rm QCD}$ given in Table 5 can accurately be represented by simple empirical formulae. We find that our central values (third column of Table 5) and error estimates due to higher order corrections can be conveniently expressed as

$$\delta_{\rm QCD} = -2.8599\ a(\xi M_t)\ , \tag{49}$$

$$\xi = 0.260^{+0.079}_{-0.056}\ , \tag{50}$$

while Eq.(36) (second column of Table 5) corresponds to $\xi = 0.339$. We emphasize that Eqs.(47,48) are not the result of a FAC optimization. They are simply empirical formulae that reproduce the values in the tables with errors of at most $1 \times 10^{-4}$ for $\xi = 0.260$ and $\xi = 0.339$, and at most $2 \times 10^{-4}$ for $\xi = 0.204$.

Combining quadratically the theoretical error in Eq.(48) with that arising from the $\pm 0.006$ uncertainty is $\alpha_s$, we obtain an overall uncertainty in $\delta_{\rm QCD}$ of $\pm(8.0,\ 8.2,\ 8.6,\ 9.7) \times 10^{-3}$ for $M_t = (220,\ 200,\ 175,\ 130)$GeV. The effect on the predicted values of $m_W$ and $\sin^2 \hat\theta_W(m_Z)$ due to a shift in $\delta_{\rm QCD}$ can be obtained using Eqs.(37a) and (37b) of Ref.[27]. We have

$$\frac{\delta m_W}{m_W} = \frac{c^2 x_t \delta(\delta_{\rm QCD})}{2(c^2 - s^2 - 2c^2 x_t)}\ , \tag{51}$$

$$\delta \hat s^2 = -\frac{3}{16\pi} \frac{\hat\alpha}{\hat c^2 - \hat s^2} \frac{M_t^2}{m_Z^2} \delta(\delta_{\rm QCD})\ , \tag{52}$$

where $x_t = 3G_\mu M_t^2/(8\sqrt{2}\pi^2)$, $s^2 = 1 - c^2 \equiv 1 - m_W^2/m_Z^2$, and $\hat{s}^2 = 1 - \hat{c}^2 \equiv \sin^2\hat{\theta}_W(m_Z)$ and $\hat{\alpha} = \hat{\alpha}(m_Z) \approx 1/127.9$ are the $\overline{\rm MS}$ parameters. For the above $M_t$ entries and $M_H = 300$GeV, and using the overall uncertainty $\delta(\delta_{\rm QCD})$, we obtain $\delta m_W = \pm(7.1,\ 6.0,\ 4.8,\ 3.0)$ MeV and $\delta\hat{s}^2 = \pm(4.0,\ 3.4,\ 2.8,\ 1.7)\times 10^{-5}$, with a very similar shift for the effective electroweak parameter $\sin^2\theta_W^{eff}$. The effect on the $M_t$ prediction from precision electroweak physics can be inferred from detailed tables such as those given in Ref.[27]. However, a simple estimate can be obtained by noting that, to a good approximation, the precision electroweak data fixes the value of $(\Delta\rho)_f$. Recalling Eq.(33), we have therefore the approximate relation

$$\delta M_t \approx -\frac{M_t}{2}\frac{\delta(\delta_{\rm QCD})}{[1+\delta_{\rm QCD}]}\ , \tag{53}$$

from which we find $\delta M_t \approx \pm(9.9,\ 9.3,\ 8.6,\ 7.2)\times 10^{-1}$ GeV.

It is interesting to note that: i) The results of the $a(M_t)$ expansion (Eq.(36)) lie almost exactly at the lower boundary (in absolute value) of the band derived by our central values (third column of Table 5) and error estimates (Eqs.(48–50)). ii) On the other hand, the values from the optimized versions of Eq.(36), namely Eqs.(40–42), are not only close among themselves, but also lie near the central values reported in this paper. iii) The latter represent a fractional enhancement of (24–22)%, depending on $M_t$, over the one–loop Djouadi–Verzegnassi result. iv) Our value for $M_t = 200$ Gev, namely $\delta_{\rm QCD} = -0.1174 \pm 0.0049$, is quite close to the estimate $\delta_{\rm QCD} = -0.121 \pm 0.006$ given in Ref.[25] before the complete $\mathcal{O}(\alpha\alpha_s^2)$ result became available.

## 3 Aknowledgments.

We would like to thank M. Beneke, K.G. Chetyrkin, J.M. Cornwall, P. Gambino, J.H. Kühn, J. Papavassiliou, G. Passarino, M. Porrati and G. Weiglein for very useful discussions. This research was supported in part by the National Science Foundation under grant No. PHY–9313781.

**Table 1**

The $A_n$ coefficients in Eq.(29). This expansion corresponds to the neglect of all contributions of $\mathcal{O}(b^l {\alpha_s}^n)$ $(l \leq n-2)$. See text.

| $n$ | 0 | 1 | 2 | 3 | 4 | 5 | 6 | 7 | 8 | 9 | 10 |
|---|---|---|---|---|---|---|---|---|---|---|---|
| $A_n$ | 1 | 2.34308 | 2.20283 | 2.28873 | 2.27584 | 2.30200 | 2.29632 | 2.30185 | 2.30003 | 2.30127 | 2.30077 |

**Table 2**

Alternative prescriptions for the expansion coefficients in Eq.(29). i) The $P_n$ replace the $A_n$ when the $1/b$ terms in the PT self–energy are incorporated. ii) The $L_n$ coefficients are similarly based on the transverse or Landau ($\xi_Q = 0$) BFG. iii) The $B_n^*$ correspond to the $\xi_Q^*$ prescription. iv) The $V_n$ correspond to the $\xi_Q = -3$ BFG. Among the BFG coefficients, the $V_n$ have the smallest asymptotic limit. These alternative prescriptions are $n_f$−dependent and are given here for the top quark case ($b = 23/3$).

| $n$ | 0 | 1 | 2 | 3 | 4 | 5 | 6 | 7 | 8 | 9 | 10 |
|---|---|---|---|---|---|---|---|---|---|---|---|
| $P_n$ | 1 | 2.78329 | 2.84487 | 2.93323 | 2.95309 | 2.98043 | 2.98150 | 2.98642 | 2.98592 | 2.98687 | 2.98664 |
| $L_n$ | 1 | 2.29111 | 2.02192 | 2.11758 | 2.09633 | 2.12387 | 2.11601 | 2.12210 | 2.11980 | 2.12122 | 2.12060 |
| $B_n^*$ | 1 | 2.13324 | 1.79728 | 1.91085 | 1.87524 | 1.90510 | 1.89496 | 1.90208 | 1.89907 | 1.90076 | 1.89998 |
| $V_n$ | 1 | 1.65829 | 1.23658 | 1.41819 | 1.32999 | 1.38095 | 1.35691 | 1.36964 | 1.36347 | 1.36663 | 1.36507 |

**Table 3**

Comparison of different evaluations of $M_t/\hat{m}_t(M_t)$. The second column gives the results of Eqs.(1,2), while the third one is based on the BLM optimization Eq.(3). The fourth and fifth columns are based on Eq.(29) with the summation truncated at $n = 6$ and $n = 7$, respectively, and the second order coefficient replaced by its exact value from Eqs.(1,2).

| $M_t/\hat{m}_t(M_t)$ | | | | |
|---|---|---|---|---|
| $M_t$ $(GeV)$ | $Eqs.(1,2)$ | $Eq.(3)$ | $Eq.(29)$ $n=6$ | $Eq.(29)$ $n=7$ |
| 130 | 1.06140 | 1.06875 | 1.06902 | 1.06951 |
| 150 | 1.05989 | 1.06673 | 1.06693 | 1.06735 |
| 175 | 1.05835 | 1.06470 | 1.06482 | 1.06518 |
| 200 | 1.05708 | 1.06303 | 1.06311 | 1.06342 |
| 220 | 1.05621 | 1.06190 | 1.06195 | 1.06222 |

**Table 4**

The correction $\Delta_{\rm QCD}$ (Eq.(44))

| $M_t$ $(GeV)$ | $10^3\Delta_{\rm QCD}$ |
|---|---|
| 130 | $-2.27$ |
| 150 | $-2.39$ |
| 175 | $-2.50$ |
| 200 | $-2.59$ |
| 220 | $-2.64$ |

**Table 5**

Comparison of several evaluations of $\delta_{\rm QCD}$.

| | $\delta_{\rm QCD}$ | | | |
|---|---|---|---|---|
| $M_t$ $(GeV)$ | $Eq.(36)$ | $Eqs.(3,43)$ | $Eqs.(29,43)$ $n=6$ | $Eqs.(29,43)$ $n=7$ |
| 130 | $-0.1205$ | $-0.1265$ | $-0.1269$ | $-0.1277$ |
| 150 | $-0.1177$ | $-0.1233$ | $-0.1236$ | $-0.1243$ |
| 175 | $-0.1149$ | $-0.1200$ | $-0.1202$ | $-0.1208$ |
| 200 | $-0.1125$ | $-0.1174$ | $-0.1175$ | $-0.1180$ |
| 220 | $-0.1109$ | $-0.1155$ | $-0.1156$ | $-0.1161$ |